\documentclass[aip,reprint]{revtex4-1}
\usepackage{graphicx}
\usepackage{dcolumn}
\usepackage{bm}
\usepackage{amsmath}%
\usepackage{amsfonts}%
\usepackage{amssymb}
\begin{document}
\title{Comparative analysis of resonant phonon THz quantum cascade lasers}
\author{Christian Jirauschek}
\email{jirauschek@tum.de}
\affiliation{Institute for Nanoelectronics, Technische Universit\"{a}%
t M\"{u}nchen, D-80333
Munich, Germany}
\author{Giuseppe Scarpa}
\affiliation{Institute for Nanoelectronics, Technische Universit\"{a}%
t M\"{u}nchen, D-80333
Munich, Germany}
\author{Paolo Lugli}
\affiliation{Institute for Nanoelectronics, Technische Universit\"{a}%
t M\"{u}nchen, D-80333
Munich, Germany}
\author{Miriam S. Vitiello}
\affiliation
{CNR-INFM Regional Laboratory LIT$^3$ and Dipartimento Interateneo di Fisica "M. Merlin", Università degli Studi di Bari, Via Amendola 173, 70126 Bari, Italy}
\author{Gaetano Scamarcio}
\affiliation
{CNR-INFM Regional Laboratory LIT$^3$ and Dipartimento Interateneo di Fisica "M. Merlin", Università degli Studi di Bari, Via Amendola 173, 70126 Bari, Italy}
\date{04 November 2009, published as J. Appl. Phys. 101, 086109 (2007)}
\begin{abstract}
We present a comparative analysis of a set of GaAs-based THz quantum cascade lasers, based on longitudinal-optical phonon scattering depopulation, by using an ensemble Monte Carlo simulation, including both carrier-carrier and carrier-phonon scattering. The simulation shows that the parasitic injection into the states below the upper laser level limits the injection efficiency and thus the device performance at the lasing threshold. Additional detrimental effects playing an important role are identified. The simulation results are in reasonable agreement with the experimental findings.
\end{abstract}
\maketitle

Quantum cascade lasers (QCLs) have an enormous potential as compact and
efficient THz sources, promising mW-level output power. Recently a new scheme
based on longitudinal-optical (LO) phonon scattering depopulation has been
successfully demonstrated,\cite{1,2,3,4} designed for high operating
temperatures and reaching up to $164\,\mathrm{K}$ in pulsed and
$117\,\mathrm{K}$ in continuous wave (cw) operation.\cite{4} In these
structures, the collector state is separated from the next lower level by at
least the LO phonon energy, enabling efficient depletion of the lower
radiative state by LO phonon scattering and reducing thermal backfilling. This
overcomes the limitations associated with the conventional bound-to-continuum
and chirped superlattice designs, where thermal backfilling of the lower
radiative state prevents high temperature performance.

In this letter, we theoretically investigate three GaAs/Al$_{0.15}$Ga$_{0.85}%
$As THz QCL structures based on the above scheme and fabricated with
high-confinement low-loss double metal waveguides. An experimental comparison
between these structures was already carried out,\cite{3} and motivates our
theoretical comparative investigation. For the analysis, an ensemble Monte
Carlo (MC) method is used,\cite{5,6,7} which has successfully been applied to
the theoretical investigation of such structures.\cite{8,9,10} The subband
energies and wavefunctions of the structure are calculated with a
Schr\"{o}dinger-Poisson solver, coupled to the MC simulation. The essential
scattering mechanisms due to electron-electron, electron-LO phonon, and
electron-acoustic phonon interactions are considered as described in Ref. 6.
Intercarrier scattering is implemented based on the Born
approximation,\cite{11,12} also taking into account screening
effects.\cite{13} In the simulated structures, intercarrier scattering plays a
major role for the intrasubband dynamics, while carrier transport is governed
by LO phonon scattering for the temperature and doping level considered.
Non-equilibrium phonon distributions are explicitly taken into account, and
the Pauli's exclusion principle is also included. Due to the relatively low
electron densities, the influence of such mechanisms is negligible. Periodic
boundary conditions are used, i.e., electrons coming out from one side of the
device are automatically injected into the equivalent level on the opposite
side.\cite{7}
\begin{figure}[ptb]
\includegraphics{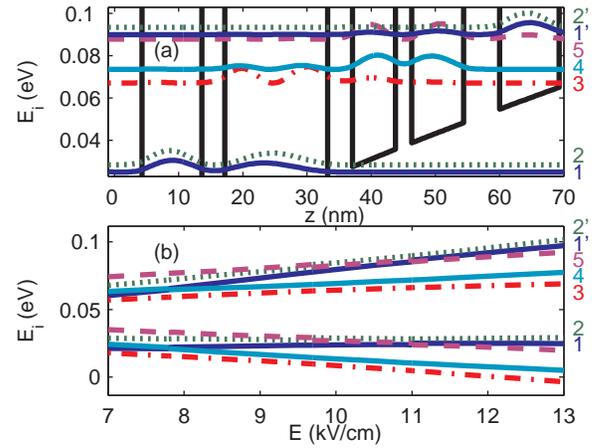}
\caption{(Color online) (a) Conduction band profile and probability densities
of a $2.8\,\mathrm{THz}$ QCL based on a phonon depopulation scheme at
$11.3\,\mathrm{kV}/\mathrm{cm}$. The layer thicknesses of one period, starting
from the injection barrier, are 56/93/36/160/39/67/25/81 [\AA]. The underlined
layer is doped at $n=1.9\times10^{16}\,\mathrm{cm}^{-3}$ and the sheet density
for one period is $n_{\mathrm{2D}}=3\times10^{10}\,\mathrm{cm}^{-2}$. (b)
Level energies as a function of the applied field for two periods of the
$2.8\,\mathrm{THz}$ structure.}%
\label{V}%
\end{figure}
Fig. 1(a) shows one period of the conduction band structure of the
$2.8\,\mathrm{THz}$ QCL\cite{3} calculated at $11.3\,\mathrm{kV}/\mathrm{cm}$;
the band profiles of the other structures discussed here are similar. The
upper and lower levels of the lasing transition around $2.8\,\mathrm{THz}$ are
labeled 5 and 4, respectively. Efficient depopulation of level 4 is achieved
via the collector state 3, which is in turn separated from the levels 2 and 1
by roughly the optical phonon energy ($36\,\mathrm{meV}$), thus enabling
efficient depletion to these levels by optical phonon scattering. Parasitic
channels counteracting the onset of population inversion between levels 5 and
4 are provided by the injection from levels 1' and 2' (i.e., the replica of
levels 1 and 2 situated in the adjacent period on the right side of Fig. 1)
into levels 1 -- 4, and by scattering processes from the upper laser level to
the states below. In Fig. 1(b), the level energies of the $2.8\,\mathrm{THz}$
design are shown as a function of the applied electric field. The strong
anticrossings between level 5 and the injection levels 2' and 1' around $9.9$
and $11\,\mathrm{kV}/\mathrm{cm}$ facilitate a selective injection into the
upper laser level, leading to a reduction of the parasitic channels. Around
$7.6\,\mathrm{kV}/\mathrm{cm}$, there is an additional anticrossing between
levels 1' and 4, causing a strong parasitic injection into the lower laser
level.\cite{8} The energy difference between the collector state 3 and the
injector state 2 (1) increases from $28\,\mathrm{meV}$ ($36\,\mathrm{meV}$)
for $7\,\mathrm{kV}/\mathrm{cm}$ to $40\,\mathrm{meV}$ ($44\,\mathrm{meV}$)
for $13\,\mathrm{kV}/\mathrm{cm}$, enabling an efficient LO-phonon-induced
depletion of the lower laser level especially for biases $>10\,\mathrm{kV}%
/\mathrm{cm}$.
\begin{figure}[ptb]
\includegraphics{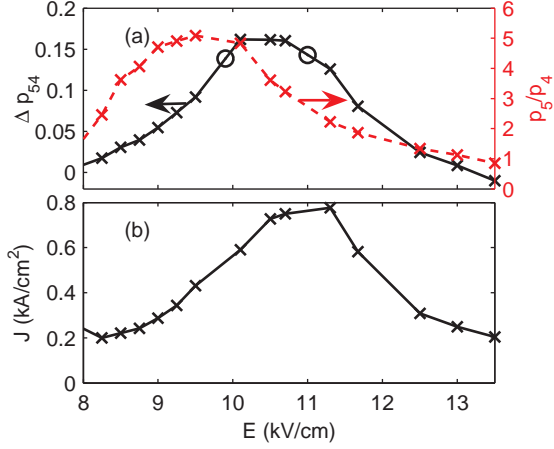}
\caption{(Color online) Simulation results for the $2.8\,\mathrm{THz}$
structure as a function of the applied field ($E$). The crosses mark the
sampling points; the lines are guide to the eye. (a) Relative population
difference (solid line) and inversion (dashed line) between the upper and
lower laser level. The circles mark the positions of the anticrossings 5 - 2'
and 5 - 1'. (b) Current density, plotted as a function of $E$.}%
\end{figure}
Figure 2 shows MC simulation results for the $2.8\,\mathrm{THz}$
structure at a lattice temperature of $100\,\mathrm{K}$. The sampling electric
fields were chosen as to avoid the excessive current spikes emerging near
narrow anticrossings in MC simulations.\cite{8} It should be pointed out that
sharp anticrossing features are an artifact of the Schr\"{o}dinger solution in
the absence of broadening mechanisms (such as those provided by surface
roughness). If a more precise quantum mechanical approach were used, based for
example on density matrices\cite{14,15,16} or nonequilibrium Green's
funcions\cite{17}, much smoother features would be observed, that would
nevertheless not drastically modify our analysis. In Fig. 2(a), the difference
and ratio of the relative population in the upper ($p_{5}$) and lower ($p_{4}%
$) laser levels is displayed, with $\Delta p_{54}=p_{5}-p_{4}$. The current
density in Fig. 2(b) exhibits a parasitic peak (not shown) around the 1' - 4
anticrossing and a local minimum at $8.25\,\mathrm{kV}/\mathrm{cm}$. It
reaches its maximum around the two anticrossings of level 5 with the injector
states 1' and 2', where the injection into subband 5 is most efficient. In
addition, the energy difference between the collector state 3 and levels 2,1
is designed to reach the LO phonon energy in this bias range and thus become
favorable for LO phonon-induced depletion of level 4. These two effects
combine to yield a large population difference $\Delta p_{54}$ and thus a high
material gain between $10$ and $11\,\mathrm{kV}/\mathrm{cm}$.

The calculated inversion peak $p_{5}/p_{4}=5.1$ lies above the experimentally
measured maximum $p_{5}/p_{4}=2.58\pm0.52$,\cite{3} and also the
experimentally measured current density peak\cite{3} is overestimated by about
40\%. This is a well known phenomenon for MC simulations of structures
dominated by LO phonon-induced carrier transport,\cite{8,18} and may partly be
due to the absence of broadening mechanisms in the algorithm, as discussed
above. To some extent, the deviation can also be ascribed to a lower than
specified free-carrier density in the sample. In addition, the experimental
effective current density, which has been calculated based on an effective
area, is smaller than the on-axis value in the gain medium. The simulated
inversion peak is shifted by $-2.4\,\mathrm{kV}/\mathrm{cm}$ relative to the
experimental maximum, which occurs at $11.9\,\mathrm{kV}/\mathrm{cm}%
$,\cite{19} and the current density peak is shifted by a similar amount.
Possible explanations are an additional parasitic resistance in the
experimental structure or deviations in the growth process.
\begin{figure}[ptb]
\includegraphics{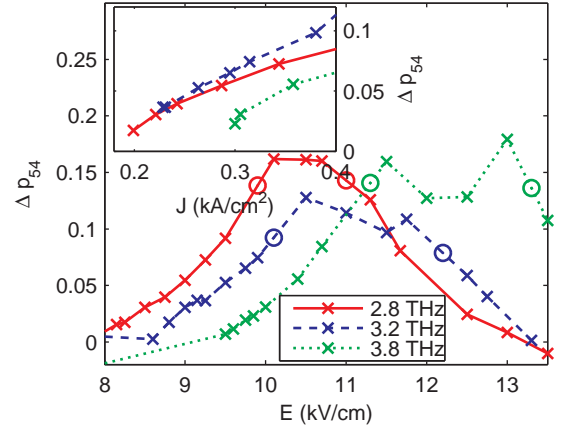}
\caption{(Color online) Population difference between upper and lower laser
level as a function of the applied field. The circles mark the positions of
the anticrossings 5 - 2' and 5 - 1'. The inset shows $\Delta p_{54}$ as a
function of the current density close to the onset of inversion. The crosses
mark the sampling points; the lines are guide to the eye.}%
\end{figure}
In Fig. 3, the population difference $\Delta p_{54}$ is plotted as
a function of the electrical field for three THz QCLs operating at
$2.8\,\mathrm{THz}$\cite{3}, $3.2\,\mathrm{THz}$\cite{2} and
$3.8\,\mathrm{THz}$\cite{1}. The maximum of $\Delta p_{54}$ in Fig. 3,
which occurs at an especially high bias for the $3.8\,\mathrm{THz}$ structure
in agreement with experiment,\cite{1} is for all designs located around the
two anticrossings of level 5 with the injector states. The inset shows $\Delta
p_{54}$ close to the onset of inversion as a function of the current density,
for values ranging between the minimum value after the parasitic 1' -- 4
anticrossing peak and $0.4\,\mathrm{kA}/\mathrm{cm}^{2}$. This corresponds to
an electrical field range of $8.25-9.4$, $9.15-10.4$ and
$9.85-10.5\,\mathrm{kV}/\mathrm{cm}$ for the $2.8$, $3.2$ and
$3.8\,\mathrm{THz}$ design, respectively. In the $2.8$ and $3.2\,\mathrm{THz}$
structures, the onset of inversion occurs at much smaller current densities
than for the $3.8\,\mathrm{THz}$ one, which is an indication of the weaker
parasitic channel 1' - 4. These findings are in agreement with the
experimental results, which yield a significantly reduced threshold current
for the $2.8$ and $3.2\,\mathrm{THz}$ structures as compared to the
$3.8\,\mathrm{THz}$ design.\cite{3}
\begin{table}[h]
\caption{Overview over simulation results for the $2.8/3.2/3.8\,\mathrm{THz}$
structure close to the onset of inversion.}
\begin{center}%
\begin{tabular}
[c]{lll}%
Quantity & $J=0.3\,$kA/cm$^{2}$ & $J=0.4\,$kA/cm$^{2}$\\
$p_{5}$ (\%) & \multicolumn{1}{c}{7.5/8.2/3.5} &
\multicolumn{1}{c}{10.6/13.8/7.9}\\
$p_{4}$ (\%) & \multicolumn{1}{c}{1.6/1.5/1.2} &
\multicolumn{1}{c}{2.1/2.5/1.3}\\
$\eta_{inj,5}$ (\%) & \multicolumn{1}{c}{43.3/42.8/26.1} &
\multicolumn{1}{c}{42.9/48.4/40.1}\\
$\eta_{inj,4}$ (\%) & \multicolumn{1}{c}{14.6/15.2/22.6} &
\multicolumn{1}{c}{16.3/18.9/13.6}\\
$\eta_{inj,3}$ (\%) & \multicolumn{1}{c}{20.9/23.6/17.8} &
\multicolumn{1}{c}{23.0/21.9/26.1}\\
$\eta_{inj,2}$ (\%) & \multicolumn{1}{c}{3.5/3.2/12.2} &
\multicolumn{1}{c}{3.4/2.2/9.3}\\
$\eta_{inj,1}$ (\%) & \multicolumn{1}{c}{14.2/10.7/19.2} &
\multicolumn{1}{c}{13.0/7.7/9.9}\\
$\tau_{5}$ (ps) & \multicolumn{1}{c}{2.75/3.05/2.02} &
\multicolumn{1}{c}{2.96/3.43/2.18}\\
$\tau_{5\rightarrow4}$ (ps) & \multicolumn{1}{c}{11.1/11.7/13.7} &
\multicolumn{1}{c}{10.1/9.7/11.9}\\
$\tau_{4}$ (ps) & \multicolumn{1}{c}{0.97/0.88/0.68} &
\multicolumn{1}{c}{0.88/0.83/0.71}\\
$f_{54}$ & \multicolumn{1}{c}{0.65/0.53/0.78} &
\multicolumn{1}{c}{0.52/0.49/0.72}%
\end{tabular}
\end{center}
\end{table}
To further investigate the structures, we introduce the injection efficiency
$\eta_{inj,i}$ from a given period into a level $i$ of the next-lower period,
defined as the ratio between the carriers injected into level $i$ and the
total amount of carriers injected into that period. An overview of the
simulation results is given in Table I. The low inversion in the
$3.8\,\mathrm{THz}$ structure at $J=0.3\,\mathrm{kA}/\mathrm{cm}^{2}$ can be
partly attributed to the poor injection efficiency into the upper laser level
$\eta_{inj,5}$, which is only 26\% for the $3.8\,\mathrm{THz}$ design, as
compared to over 40\% for the other structures. This is due to the strong
parasitic injection into the states below the upper laser level, as indicated
by the increased $\eta_{inj,1..4}$. From Table I we can see that also for the
$2.8$ and $3.2\,\mathrm{THz}$ designs, the parasitic injection channels to
levels 1 -- 4 play a significant role in the threshold region. To obtain a
high inversion, long lifetimes $\tau_{5}$ and $\tau_{5\rightarrow4}$ are
desirable,\cite{20} as well as a small $\tau_{4}$ ensuring a fast depopulation
of the lower laser level. Furthermore, the gain is directly proportional to
the reduced oscillator strength $f_{54}$.\cite{8} $\tau_{5}$ in the
$3.8\,\mathrm{THz}$ structure is reduced mainly due to an increased leakage to
levels 3 and 1, while the values of $\tau_{4}$ and $f_{54}$ are slightly
superior for this design. At low current densities, the $2.8\,\mathrm{THz}$
structure shows a higher value of $f_{54}$ than the $3.2\,\mathrm{THz}$ laser,
outweighing the somewhat lower inversion and resulting in a slightly better
threshold performance, as experimentally observed.\cite{3}

We have also calculated the electron-lattice energy relaxation rate $\tau
_{E}^{-1}$.\cite{21} Such quantity reaches its peak in the lasing region,
where the phonon-assisted injection into the upper laser level and depletion
of the lower laser level are the strongest. For the best of the structures
($2.8\,\mathrm{THz}$), a maximum $\tau_{E}^{-1}$ of approximately
$3.8\,\mathrm{ps}^{-1}$ was calculated, in reasonable agreement with
experiment.\cite{3} The other two structures show a lower maximum value of
$\tau_{E}^{-1}$ ($3.3\,\mathrm{ps}^{-1}$), also in qualitative agreement with
the experiment. For some of the subbands, the simulation yields highly
nonequilibrium carrier distributions, complicating comparison with
experimentally extracted temperatures.\cite{3} However, an investigation of
this issue goes beyond the framework of this paper and will be addressed in a
different article.

In conclusion, we have shown that the MC simulation allows us to identify the
parasitic processes affecting the operation of THz lasers, thus providing a
meaningful explanation of the experimental findings.

We would like to acknowledge the support of this research by the EOARD Grant
No. 063008 and by the Emmy Noether program of the Deutsche
Forschungsgemeinschaft. CNR-INFM LIT3 Laboratory acknowledges partial
financial support from MIUR Project No. FIRB-RBAU01E8SS and DD1105/2002.


\begin{thebibliography}{9}
\bibitem {19}The experimental electric field values were computed from the
measured voltages subtracting the voltage drop at the top Schottky barrier and
dividing the resultant bias by the active region thickness, assuming a
constant field in the gain medium.
\bibitem {20}$\tau_{5}$ and $\tau_{5\rightarrow4}$ are defined as the
characteristic times for relaxation from level 5 to all lower levels and to
level 4 alone, respectively, and already include the effect of backscattering
from the final state.\cite{8}
\bibitem {21}$\tau_{E}$ is here calculated as $\tau_{E}=\left(
E_{\mathrm{kin},e}-E_{\mathrm{kin},e}^{\mathrm{eq}}\right)  /P_{\mathrm{th}}$,
where $E_{\mathrm{kin},e}$ is the total kinetic energy of the $N$ simulated
electrons and $E_{\mathrm{kin},e}^{\mathrm{eq}}=Nk_{\mathrm{B}}T_{\mathrm{L}}$
is the equilibrium kinetic energy at lattice temperature $T_{\mathrm{L}}$. The
dissipated thermal power $P_{\mathrm{th}}$ is computed as the product of the
calculated current and the applied voltage.
\end{thebibliography}
\end{document}